\documentclass[letter,twocolumn]{jpsj2}
\usepackage{txfonts}
\usepackage{bm}

\title{Theory of Orbital Susceptibility on Excitonic Insulator  }

\author{
Hiroyasu Matsuura\thanks{matsuura@hosi.phys.s.u-tokyo.ac.jp} and Masao Ogata}

\inst{
Department of Physics, University of Tokyo, Hongo, Bunkyo-ku, Tokyo 113-0033, Japan \\
} 

\abst{We study the temperature dependence of the orbital susceptibility of an excitonic insulator on the basis of a two-band model.
It is shown that a drastic change (an anomalous enhancement) in susceptibility as a function of temperature occurs owing to the occurrence of additional orbital susceptibility due to the excitonic gap. 
We calculate explicitly the temperature dependence of orbital susceptibility for a model of Ta$_2$NiSe$_5$, and show that the result is consistent with experimental results.
}

\begin{document}
\maketitle
In semimetals, electrons and holes form bound pairs of excitons.
These bound pairs lead to a nonconducting state called excitonic insulator.
In the 60s much interest was attracted by the excitonic insulator~\cite{Mott,Kozlov,Jerome,Zittartz1,Zittartz2,Zittartz3,Zittartz4,Fenton,Halperin}.
 One of the important and interesting points concerning the excitonic insulator is that the excitonic theory is transformed to the BCS theory by a particle-hole transformation.  
From the similarity to the BCS theory, the stability of the excitonic insulator~\cite{Kozlov,Jerome,Halperin}, anisotropy~\cite{Zittartz1}, impurity effect~\cite{Zittartz2}, transport properties~\cite{Jerome,Zittartz3,Zittartz4}, and effect of magnetic field~\cite{Fenton} have been extensively studied. 
However, no materials have been identified as an excitonic insulator.

Recently, it has been argued that Ta$_2$NiSe$_5$ is a candidate material of an excitonic insulator from photoemission spectroscopy~\cite{Wakisaka,Seki}.
In this experiment, the observed energy dispersion seems to be consistent with the theoretical result based on the mean field approximation of a model consisting of electron and hole bands with Coulomb interactions between them.
Since this experiment, the stability of the excitonic insulator has been studied on the basis of numerical calculations of the two-band model or an extended model~\cite{Kaneko1,Kaneko2,Kaneko3,Yamada}.

Several experiments have been suggested from theoretical viewpoints using the relationship between the excitonic insulator and superconductivity to confirm the existence of the excitonic insulator~\cite{Sugimoto}.
For example, the temperature dependences of the ultrasonic absorption coefficient and nuclear magnetic relaxation rate have been calculated~\cite{Sugimoto}.
However, there have been no experimental results that confirm the excitonic insulator except for photoemission spectroscopy.  
It is well known that Ta$_2$NiSe$_5$ shows a drastic change in magnetic susceptibility at the transition temperature~\cite{Salvo}.  
Although this temperature dependence is expected to be a feature of the excitonic insulator, the origin of this drastic change has not been understood theoretically.
In this Letter, we study the temperature dependence of orbital susceptibility in a two-band model for the excitonic insulator.
We clarify that the temperature dependence of susceptibility is understood as an occurrence of additional orbital susceptibility owing to the excitonic phase transition. 
We show that the obtained temperature dependence of orbital susceptibility  agrees very well with experiments.

\begin{figure}
\rotatebox{0}{\includegraphics[angle=-90,width=1\linewidth]{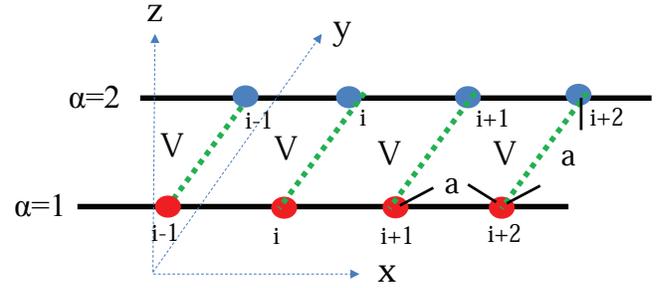}}
\caption{(Color online) Schematic picture of two-chain model.
A magnetic field is applied to the z-axis.
Red and blue circles indicate the lattice sites, $\alpha$ being the index of the chain.
$V$ is the Coulomb interaction between two chains.
}
\label{Fig1}
\end{figure}

To study the orbital susceptibility of the excitonic insulator, we use a two-band model consisting of electron and hole bands with Coulomb interactions under a magnetic field.
We use a recently developed exact formalism for orbital susceptibility~\cite{OgataFukuyama,Ogata,Matsuura}.
A schematic picture of this model is shown in Fig.\ \ref{Fig1}, where $\alpha$ is the index of two chains $\alpha=1,2$, and $a$ is the lattice constant between two sites.
Red and blue circles indicate the lattice sites, and $V$ indicates the Coulomb interaction between chains.

First, we construct a Hamiltonian of the effective model of Fig.\ \ref{Fig1}.
A one-body Hamiltonian under the magnetic field is given by
\begin{eqnarray}
\hat{H}({\bf r}) = \frac{1}{2m}\bigr({\bf p} -\frac{e}{c}{\bf A}({\bf r})\bigr)^2 + \sum_{\alpha,j}V({\bf r}_{\alpha,j}), \label{Hamiltonian}
\end{eqnarray}
where $m$ and $e$ ($e<0$) are the mass and charge of an electron, respectively. $c$ is the speed of light, ${\bf A}({\bf r})$ is a vector potential, and $V({\bf r}_{\alpha,j})$ is a potential at ${\bf r}_{\alpha,j}= {\bf r}-{\bf R}_{\alpha,j}$, where ${\bf R}_{\alpha,j}$ is the position of the $j$-th site of the $\alpha$ chain.

The wave function under the magnetic field is obtained as~\cite{Matsuura}
\begin{eqnarray}
\Phi({\bf r}_{\alpha,i})= \exp{\bigr( \frac{ie}{c\hbar}{\bf A}_{\alpha,i} \cdot {\bf r} \bigr)} \phi({\bf r}_{\alpha,i}),  \label{WF}
\end{eqnarray}
where ${\bf A}_{\alpha,i} = {\bf A}({\bf R}_{\alpha,i})$.
$\phi({\bf r}_{\alpha,i})$ is an eigenstate of
\begin{eqnarray}
\bigr[ \frac{1}{2m}p^2 + V({\bf r}_{\alpha,i}) \bigr]\phi({\bf r}_{\alpha,i})= \epsilon_0\phi({\bf r}_{\alpha,i}).
\end{eqnarray}
where $\epsilon_0$ is an eigenvalue.
We assume the orthogonality of the wave functions 
\begin{eqnarray}
\int {\rm d}{\bf r}\Phi^{*}({\bf r}_{\alpha,i})\Phi({\bf r}_{\alpha^{\prime},j}) = \delta_{ij}\delta_{\alpha\alpha^{\prime}},
\end{eqnarray}
for simplicity.

The matrix element of Eq.\ (\ref{Hamiltonian}) is obtained as
\begin{eqnarray}
&& \int {\rm d}{\bf r}\Phi^{*}({\bf r}_{\alpha,i}) \hat{H}({\bf r})\Phi({\bf r}_{\alpha^{\prime},j}) \nonumber \\
 && = e^{-i\Phi_{ij}^{\alpha\alpha^{\prime}}}\int {\rm d}{\bf r}e^{i\chi_{ij}^{\alpha\alpha^{\prime}}} \phi^{*}({\bf r}_{k,i}) \tilde{H}({\bf r})\phi({\bf r}_{\alpha^{\prime},j}), \label{matrix0}
\end{eqnarray}
where 
\begin{eqnarray}
\tilde{H}({\bf r}) = \frac{1}{2m}\bigr({\bf p} -\frac{e}{c}({\bf A}({\bf r})-{\bf A}_{\alpha^{\prime},j})\bigr)^2 + \sum_{\alpha,j}V({\bf r}_{\alpha,j}).
\end{eqnarray}
The phases $\Phi_{ij}^{\alpha\alpha^{\prime}}$ and $\chi_{ij}^{\alpha\alpha^{\prime}}$ are given by
\begin{eqnarray}
\Phi_{ij}^{\alpha\alpha^{\prime}} &=& \frac{e}{c\hbar}({\bf A}_{\alpha,i} -{\bf A}_{\alpha^{\prime},j})\cdot \frac{{\bf R}_{\alpha,i} + {\bf R}_{\alpha^{\prime},j}}{2}, \label{PP}\\
\chi_{ij}^{\alpha\alpha^{\prime}} &=& -\frac{e}{c\hbar}({\bf A}_{\alpha,i} -{\bf A}_{\alpha^\prime,j} )\cdot ({\bf r}-\frac{{\bf R}_{\alpha,i} + {\bf R}_{\alpha^{\prime},j}}{2}).
\end{eqnarray}
Here, $\Phi_{ij}^{\alpha\alpha^{\prime}}$ and $\chi_{ij}^{\alpha\alpha^{\prime}}$ are the Peierls phase and additional phase, respectively~\cite{Matsuura}.
In order to study the magnetic field along the $y$- and $z$-axes, we use Landau gauges as follows.
\begin{eqnarray}
{\bf A}({\bf r})=\left\{ \begin{array}{ll}
(Hz,0,0) & {\rm for\  {\it y} \ direction},  \\
(0,Hx,0) & {\rm  for\ {\it z} \ direction},  \\
\end{array} \right. \label{Landau0}
\end{eqnarray} 
where $H$ is the magnetic field. 
By substituting the Landau gauges of Eq.\ (\ref{Landau0}) into Eq.\ (\ref{PP}),  the Peierls phases are obtained as
\begin{eqnarray}
\Phi_{ij}^{\alpha\alpha^{\prime}}=\left\{ \begin{array}{ll}
0 & {\rm for\ {\it y} \ direction}, \label{Landau1} \\
\frac{eH}{2c\hbar}(x_{\alpha,i}-x_{\alpha^{\prime},j})(y_{\alpha,i}+y_{\alpha^{\prime},j}) & {\rm  for\ {\it z} \ direction}, \label{Landau2} \\
\end{array} \right.
\end{eqnarray} 
where $x_{\alpha,i}$ and $y_{\alpha^{\prime},i}$ are the $x$- and $y$-components of ${\bf r}-{\bf R}_{\alpha,i}$, respectively.
$\chi_{ij}^{\alpha\alpha^{\prime}}$ is obtained in the same way.

Since the general formula of the matrix elements of Eq.\ (\ref{matrix0}) is very complicated~\cite{Matsuura}, in this Letter we use a simple formula as follows.
For $i \neq j$, the matrix element is 
\begin{eqnarray}
{\rm eq}.\ (\ref{matrix0}) \simeq e^{-i\Phi_{ij}^{\alpha\alpha^{\prime}}}(-1)^{\alpha}\delta_{\alpha,\alpha^{\prime}}( -t_0 +  t_{2}h^2 ), \label{taransfer}
\end{eqnarray}
where $t_0$ and $t_2$ are the transfer integral and the correction of the transfer integral under the magnetic field, respectively.
The detail of $t_2$ is discussed in Ref.~\citen{Matsuura}.
$h$ is the dimensionless magnetic field defined as
\begin{eqnarray}
h = \frac{eH}{2c\hbar}a^2.
\end{eqnarray}
For $i =j$ and $\alpha=\alpha^{\prime}$, the matrix element is
\begin{eqnarray}
{\rm Eq}.\ (\ref{matrix0}) \simeq (-1)^{\alpha}\epsilon +e_2h^2 -\mu,
\end{eqnarray}
where $\epsilon$ and $\mu$ are a one-body level and a chemical potential, respectively, and $e_2$ is the coefficient of $h^2$, which comes from the term of $({\bf A}({\bf r})-{\bf A}_{\alpha,j})^2$.

Next, we study the Coulomb interaction $V$ between two chains, which is expressed as 
\begin{eqnarray}
V &\equiv& \int d{\bf r}d{\bf r}^{\prime} \Phi^{*}({\bf r}_{\alpha,i})\Phi^{*}({\bf r}^{\prime}_{\alpha^{\prime},i}) \frac{e^2}{|{\bf r}-{\bf r}^{\prime}|} \Phi({\bf r}_{\alpha,i})\Phi({\bf r}^{\prime}_{\alpha^{\prime},i}). 
\end{eqnarray}
By substituting Eq.\ (\ref{WF}), it is found that $V$ is independent of the magnetic field.

As a result, the Hamiltonian of an effective model of Fig.\ \ref{Fig1} becomes
\begin{eqnarray}
H &=& \sum_{i}\bigr[( -t(h)c_{1,i}^{\dagger}c_{1,i+1} +t(h)c_{2,i}^{\dagger}c_{2,i+1} + {\rm h.c.} ) \nonumber \\
 &&+ (-\epsilon-\mu+ e_2h^2)n_{1,i} + (\epsilon-\mu+ e_2h^2)n_{2,i} \bigr] \nonumber \\
&&+ V\sum_{i}n_{1,i}n_{2,i}.
\end{eqnarray}
$c_{\alpha,i}$ is an annihilation operator at the i-site of chain $\alpha$.
The spin degrees of freedom are neglected here.
The effect of the orbital magnetic field is included in the transfer integrals $t(h) \equiv e^{-i\Phi_{i,i+1}}( -t_0 +  t_{2}h^2 )$ .

Using a mean field approximation as 
\begin{eqnarray}
Vn_{1,i}n_{2,i} &\rightarrow& -V\langle c_{1,i}^{\dagger}c_{2,i} \rangle c_{2,i}^{\dagger}c_{1,i}-V\langle c_{2,i}^{\dagger}c_{1,i} \rangle c_{1,i}^{\dagger}c_{2,i} \nonumber \\
&& +V\langle c_{1,i}^{\dagger}c_{2,i} \rangle \langle c_{2,i}^{\dagger}c_{1,i} \rangle,
\end{eqnarray}
we define an order parameter of the excitonic insulator as
\begin{eqnarray}
\Delta_{{\rm EX}} = -V\langle c^{\dagger}_{1}c_{2} \rangle = -\frac{V}{N}\sum_{i}\langle c_{1,i}^{\dagger}c_{2,i} \rangle,
\end{eqnarray}
where $N$ is the total number of sites.
We assume that the order parameter is independent of the sites.
Then, the mean-field Hamiltonian is given by
\begin{eqnarray}
H_{{\rm tot}} &=& \sum_{i}\bigr[( -t(h)c_{1,i}^{\dagger}c_{1,i+1} +t(h)c_{2,i}^{\dagger}c_{2,i+1} + {\rm h.c.} ) \nonumber \\
&&+ (-\epsilon-\mu+e_2h^2)n_{1,i} + (\epsilon-\mu+e_2h^2)n_{2,i} \nonumber \\
&& + \sum_{i}\Delta_{{\rm EX}}  c_{2,i}^{\dagger}c_{1,i}+\Delta_{{\rm EX}} c_{1,i}^{\dagger}c_{2,i} + \frac{N\Delta_{{\rm EX}}^{2}}{V} . \label{MF_Hami}
\end{eqnarray}

First, we study the case without a magnetic field.
In this case, the transfer integral is $t(h=0) = -t_0$.
By Fourier transformation, the Hamiltonian (\ref{MF_Hami}) is transformed to
\begin{eqnarray}
H_{{\rm tot}} &=& \sum_{k}\bigr[ (-\xi_{k}-\mu)c_{1,k}^{\dagger}c_{1,k} +(\xi_{k}-\mu)c_{2,k}^{\dagger}c_{2,k} +\Delta_{{\rm EX}}c_{1,k}^{\dagger}c_{2,k} \nonumber \\ 
&& + \Delta_{{\rm EX}}c_{2,k}^{\dagger}c_{1,k} \bigr] + \frac{N\Delta_{{\rm EX}}^{2}}{V},  \label{MF_eq}
\end{eqnarray}
where $\xi_k = -2t_0\cos{k_xa}+\epsilon $.
By diagonalizing this Hamiltonian, we obtain 
\begin{eqnarray}
H_{{\rm tot}} &=& \sum_{k,\pm} E_{k,\pm}p_{k\pm}^{\dagger}p_{k\pm} + \frac{N\Delta_{{\rm EX}}^{2}}{V},
\end{eqnarray}
where $p_{k\pm}$ is an annihilation operator of a quasiparticle, and the quasiparticle energy is given by
\begin{eqnarray}
E_{k,\pm} = -\mu \pm \sqrt{\xi_k^2 + \Delta_{{\rm EX}}^{2}}. 
\end{eqnarray}
In the following, we discuss only the half-filling case.
Then, the chemical potential is always $\mu=0$. 
The gap equation becomes the usual form as
\begin{eqnarray}
\Delta_{{\rm EX}} = \frac{V\Delta_{{\rm EX}}}{2N}\sum_{k}\frac{\tanh{\frac{\beta E_{k,0}}{2}}}{E_{k,0}},
\end{eqnarray}
where $E_{k,0}= \sqrt{\xi_k^2 + \Delta_{{\rm EX}}^{2}}$.
By solving this gap equation self-consistently, we obtain the amplitude and temperature dependence of the order parameter.

\begin{figure}
\rotatebox{0}{\includegraphics[angle=0,width=1\linewidth]{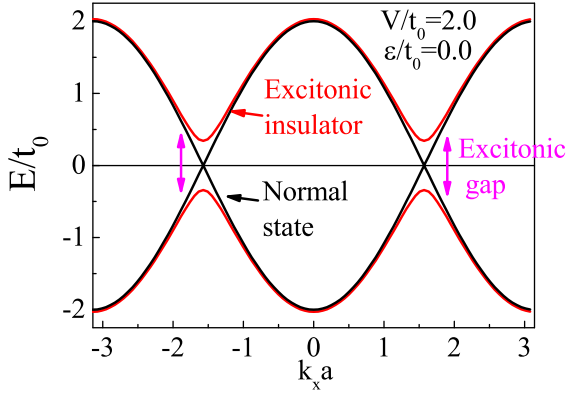}}
\caption{(Color online) Dispersion of normal state (black line) and excitonic insulator (red line). 
The parameters are set as $\epsilon/t_0 = 0.0$ and $V/t_0=2.0$.}
\label{Fig2}
\end{figure}

\begin{figure}
\rotatebox{0}{\includegraphics[angle=0,width=1\linewidth]{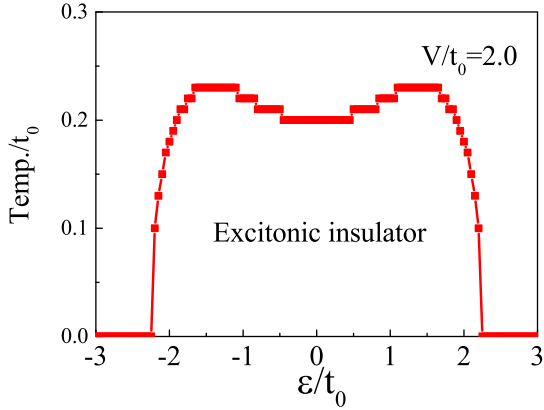}}
\caption{(Color online) $\epsilon$ dependence on the transition temperature at $V/t_0=2.0$. }
\label{Fig3}
\end{figure}

To show the feature of the excitonic insulator, we show in Fig. \ref{Fig2} the dispersions of the normal state (black line) and the excitonic insulator (red line) at $\epsilon/t_0 = 0.0$ and $V/t_0=2.0$ at half-filling.
It is found that an excitonic gap opens at $k_xa= \pm \pi/2$. 
The $\epsilon/t_0$ dependence of the transition temperature is shown in Fig. \ref{Fig3}.
It is also found that the transition temperature becomes maximum near the band edge.

Next, we calculate the orbital susceptibility in this system.
As discussed in Eq.\ (\ref{taransfer}), the transfer integral is modified by the magnetic field along the $y$-axis as $t(h)=-t_0 +t_2h^2$, while the transfer integral modified by that along the $z$-axis includes the Peierls phase.
The effective Hamiltonian under the magnetic field along the y-axis is given by replacing the $\xi_k$ of Eq.\ (\ref{MF_eq}) with $\zeta_k= 2(-t_0 +t_2h^2)\cos{k_xa} -\epsilon$.
As a result, the eigenvalues are obtained as
\begin{eqnarray}
E_{k,\pm} = -\mu + e_2h^2 \pm \sqrt{\zeta_k^2 + \Delta_{\rm EX}}.
\end{eqnarray}
In a similar way, under the magnetic field along the z-axis, the effective Hamiltonian is given by
\begin{eqnarray}
H_{{\rm tot}} &=& \sum_{k}\bigr[ (-\zeta_{1,k} -\mu +e_2h^2)c_{1,k}^{\dagger}c_{1,k} \nonumber \\
&&+(\zeta_{2,k}-\mu +e_2h^2)c_{2,k}^{\dagger}c_{2,k} \nonumber \\ 
&& +\Delta_{{\rm EX}}c_{1,k}^{\dagger}c_{2,k} + \Delta_{{\rm EX}}c_{2,k}^{\dagger}c_{1,k} \bigr] + \frac{N\Delta_{{\rm EX}}^{2}}{V},  \label{eff_ham_mag}
\end{eqnarray}
where $\zeta_{1,k}$ and $\zeta_{2,k}$ are obtained as
\begin{eqnarray}
\zeta_{1,k} &=& 2(-t_0 +t_2h^2)\cos{(k_x+h)} +\epsilon, \\
\zeta_{2,k} &=& 2(-t_0 +t_2h^2)\cos{(k_x+2h)} +\epsilon. 
\end{eqnarray}
The eigenvalues of Eq.\ (\ref{eff_ham_mag}) are given by
\begin{eqnarray}
E_{k,\pm} = -\mu + e_2h^2 + \frac{1}{2}\bigr[ \zeta_{2,k}-\zeta_{1,k} \pm \sqrt{(\zeta_{1,k}+\zeta_{2,k})^2 +4\Delta_{\rm EX}^2}\bigr]. 
\end{eqnarray}
The thermodynamic potential of the system under the magnetic field, $\Omega_{{\rm tot}}(H)$, is defined as
\begin{eqnarray}
e^{-\beta\Omega_{{\rm tot}}(H)} &=& {\rm Tr}\exp{(-\beta H_{{\rm tot}})}. 
\end{eqnarray}
By substituting the Hamiltonian, the thermodynamic potential is obtained as
\begin{eqnarray}
\Omega_{{\rm tot}}(h) = \frac{N\Delta_{{\rm EX}}^{2}}{V} - \frac{1}{\beta}\sum_{k,\pm}\ln{(1 + e^{-\beta E_{k,\pm}})}.
\end{eqnarray}
The orbital susceptibility is calculated from
\begin{eqnarray}
\chi = -\biggr[ \frac{1}{H}\frac{\partial \Omega_{{\rm tot}}(H)}{\partial H}\biggr]_{H=0} 
     = -\frac{e^2a^4}{4c^2\hbar^2}\biggr[ \frac{1}{h}\frac{\partial \Omega_{{\rm tot}}(h)}{\partial h}\biggr]_{h=0},
\end{eqnarray}
where the chemical potential is always fixed at $\mu=0$ because we discuss only the half-filling.

\begin{figure}
\rotatebox{0}{\includegraphics[angle=0,width=1\linewidth]{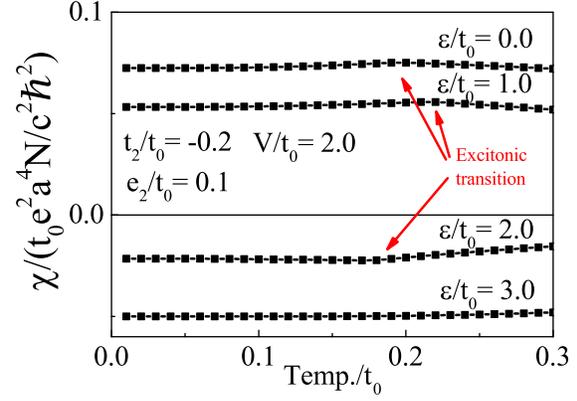}}
\caption{(Color online) Temperature dependence of the orbital susceptibility at $V/t_0=2.0$, $t_2/t_0 = -0.2$, $e_2/t_0=0.1$, and $\epsilon/t_0= 3.0$, $2.0$, $1.0$, and $0.0$.
The magnetic field is applied along the $y$-axis.
 }
\label{Dia}
\end{figure}

Figure \ref{Dia} shows the temperature dependence on the orbital susceptibility at $V/t_0=2.0$, $t_2/t_0 = -0.2$, and $\epsilon=3.0$, $2.0$, $1.0$, and $0.0$ under a magnetic field along the $y$-axis.
The red arrows indicate the transition temperatures of the excitonic insulator for each parameter.
The explicit transition temperature is shown in Fig.\ \ref{Fig3}.
Above the transition temperature, the temperature dependence of susceptibility depends on the amplitude of $\epsilon$: the susceptibility decreases for $\epsilon/t_0=0.0$ and $1.0$ and increases for $\epsilon/t_0=2.0$ and $3.0$ as the temperature increases.
Below the transition temperature, the temperature dependence of the susceptibility is almost constant and independent of the amplitude of $\epsilon$.  

\begin{figure}
\rotatebox{0}{\includegraphics[angle=0,width=1\linewidth]{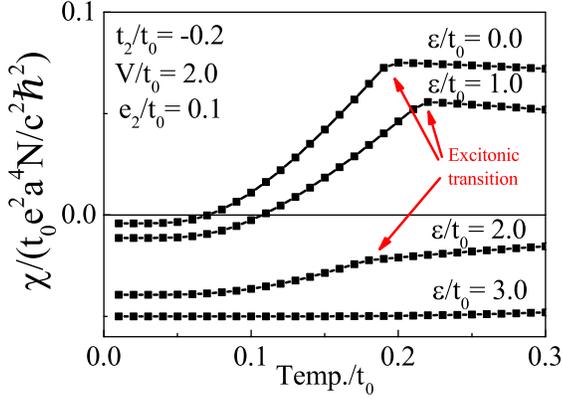}}
\caption{(Color online) Temperature dependence of the orbital susceptibility at $V/t_0=2.0$, $t_2/t_0 = -0.2$, $e_2/t_0=0.1$, and $\epsilon= 3.0$, $2.0$, $1.0$, and $0.0$. 
The magnetic field is applied to the $z$-axis.
}
\label{Dia2}
\end{figure}

Figure\ \ref{Dia2} shows the temperature dependence of the orbital susceptibility for $V/t_0=2.0$, $t_2/t_0=-0.2$, and $\epsilon/t_0=0.0$, $1.0$, $2.0$, and $3.0$ under the magnetic field along the $z$-axis.
Above the transition temperature, the temperature dependence of the susceptibility is the same as that shown in Fig.\ \ref{Dia}.
However, below the transition temperature, the susceptibility drastically decreases as the temperature decreases. 
For $\epsilon/t_0=0.0$ and $1.0$, the sign of the susceptibility changes from positive to negative as the temperature decreases.
For $\epsilon/t_0=2.0$, the susceptibility increases negatively.

In order to understand the temperature dependence on the susceptibility in detail, we divide the susceptibility into three components $\chi_{\rm PP}$, $\chi_{\rm t_2}$, and $\chi_{\rm core}$, where $\chi_{\rm PP}$, $\chi_{\rm t_2}$, and $\chi_{\rm core}$ are the susceptibilities coming from the Peierls phase, $t_2$, and $e_2$, respectively.
The total susceptibility is obtained as
\begin{eqnarray}
\chi = \chi_{\rm PP} + \chi_{\rm t_2} + \chi_{\rm core}. 
\end{eqnarray}
These susceptibilities are calculated as follows~\cite{Matsuura}:
For $\chi_{\rm PP}$, the magnetic field is applied along the $z$-axis, and the parameters are set as $V/t_0=2.0$, $t_2/t_0=0.0$, and $e_2/t_0=0.0$.
In the same way, for $\chi_{\rm t_2}$ ($\chi_{\rm core}$), the magnetic field is applied along the $y$-axis, and the parameters are set as $V/t_0=2.0$, $t_2/t_0=-0.2$, and $e_2/t_0=0.0$ ($V/t_0=2.0$, $t_2/t_0=0$, and $e_2/t_0=0.1$). 

\begin{figure}
\rotatebox{0}{\includegraphics[angle=0,width=1\linewidth]{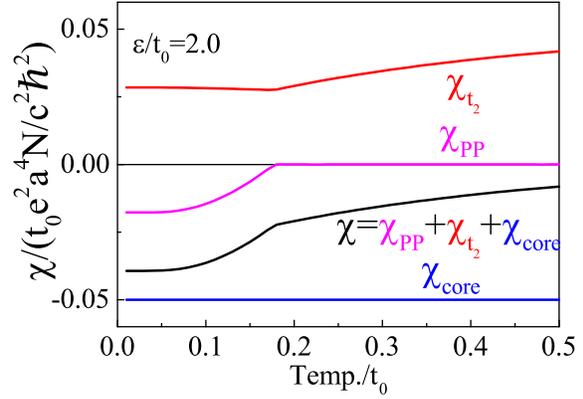}}
\caption{(Color online) Temperature dependence on the susceptibilities, $\chi_{\rm PP}$, $\chi_{t_2}$, $\chi_{{\rm core}}$, and $\chi$ at $\epsilon/t_0=2.0$.
Details of the parameter set are indicated in text.}
\label{chi3}
\end{figure}

Figure \ref{chi3} shows the temperature dependence on the susceptibilities of $\chi_{\rm PP}$, $\chi_{t_2}$, $\chi_{{\rm core}}$, and $\chi$ at $\epsilon/t_0=2.0$.
It is found that above the transition temperature, the temperature dependence on the total susceptibility comes from $\chi_{\rm t_2}$.
It is also found that below the transition temperature, the temperature dependence is derived from $\chi_{\rm pp}$.
Although the temperature dependence of $\chi_{\rm t_2}$ also changes below the transition temperature, $\chi_{\rm t_2}$ is independent of temperature in this case.
We discuss the physical meaning of the temperature dependences of $\chi_{{\rm core}}$, $\chi_{\rm t_2}$, and $\chi_{\rm pp}$.
Since $\chi_{{\rm core}}$ is the atomic diamagnetism, it is independent of the temperature, as shown in Fig.\ \ref{chi3}.
The origin of $\chi_{\rm t_2}$ is the correction of the transfer integral under the magnetic field as discussed in Eq.\ (\ref{taransfer}).
Above the transition temperature, the valence band and conduction band touch at $k_xa=\pi$ and $-\pi$.
Since the energy gain owing to the correction of the transfer integral is zero at the zero temperature of the normal state, the origin of the temperature dependence of $\chi_{\rm t_2}$ is a thermal excitation.
When the excitonic gap occurs below the transition temperature, the susceptibility due to the thermal excitation becomes smaller.
On the other hand, the mixing between the conduction band and the valence band occurs below the transition temperature.
As a result, it is expected that a finite energy gain owing to the correction of the transfer integral occurs and it is the origin of $\chi_{\rm t_2}$.    
However, the details of the temperature dependence are still unclear and this is a future research topic. 
The origin of the finite contribution of $\chi_{\rm pp}$ below the transition temperature is due to the coherent hopping between chains induced by the formation of the excitonic state.
As a result, the effect of the Peierls phase flowing between two chains increases as the temperature decreases.  

Finally, we compare the present theoretical result with experiments on Ta$_2$NiSe$_5$.
As discussed in Introduction, the temperature dependence of the susceptibility of Ta$_2$NiSe$_5$ has been reported~\cite{Salvo}.
As the temperature decreases from 900 K, the susceptibility decreases almost linearly, and the susceptibility decreases drastically below 400 K.
The susceptibility becomes independent of temperature below 200 K.  
When we set the hopping parameter as $t_0 \sim 0.2$ eV, the present result $\chi$ of Fig.\ \ref{chi3} is consistent with the temperature dependence of Ta$_2$NiSe$_5$.
Recently, the electronic state of Ta$_2$NiSe$_5$ has been studied in detail, and the effective model is known to be a three-band model~\cite{Yamada,Sugimoto}.
Although the essential point of the orbital susceptibility is understood on the basis of the two-band model in this paper, it is necessary to discuss the orbital susceptibility on the basis of the three-band model in order to understand the orbital susceptibility of Ta$_2$NiSe$_5$ in detail.   

In conclusion, we studied the temperature dependence of orbital susceptibility on the excitonic insulator on the basis of a two-band model with the magnetic field.
We clarified that the temperature dependence of the susceptibility is derived from the occurrence of additional orbital susceptibility due to the excitonic gap. 
We calculated the temperature dependence of orbital susceptibility and the result is consistent with the experimental result of Ta$_2$NiSe$_5$.

\begin{acknowledgment}
This work is supported by the JSPS Core-to-Core Program, A. Advanced Research Networks.
We are also supported by Grants-in-Aid for Scientific Research from the Japan Society for the Promotion of
Science (Nos. 15K17694, 25220803, and 15H02108).
 \end{acknowledgment}


\def\journal#1#2#3#4{#1 {\bf #2}, #3 (#4) }
\def\PR{Phys.\ Rev.}
\def\PRB{Phys.\ Rev.\ B}
\def\PRL{Phys.\ Rev.\ Lett.}
\def\JPSJ{J.\ Phys.\ Soc.\ Jpn.}
\def\PTP{Prog.\ Theor.\ Phys.}
\def\JPCS{J.\ Phys.\ Chem.\ Solids}
\def\RMP{Rev.\ Mod.\ Phys.\ }

\end{document}